\documentstyle[epsfig,epsf]{article}
\begin{document}
\vskip 2 cm
\begin{center}
{\Large {\bf Introduction to the Workshop\par 
``30 Years of Bubble Chamber Physics"}}
\end{center}

\vskip .7 cm

\begin{center}
Giorgio Giacomelli \par~\par
{\it Phys. Dept.of the University of Bologna and INFN, Sezione di Bologna} 

E-mail: giacomelli@bo.infn.it

\par~\par

\bf 18 March 2003, Bologna Academy of Sciences, Via Zamboni 31, 40126 Bologna

\vskip .7 cm
{\large \bf Abstract}\par
\end{center}

{\normalsize After some recollections of the early bubble chamber
  times, a brief overview of the golden age of the field is made,
  including its legacy and the use of bubble chamber events for the 
popularization of science.  }

\vspace{5mm}

\large
On behalf of the organizing committee I would like to welcome you to Bologna 
and to this meeting. Several colleagues sent messages apologizing for their 
absence, mainly because of health problems. We wish them well. Some of them 
sent short notes on their reminiscences of the bubble chamber golden 
age [1][2].\par
The aim of the meeting is to recall the activities connected with the bubble 
chamber technique, the main physics discoveries, the evolution of the 
international collaborations, the bubble chamber legacy, and to honour some 
of the founding fathers, in particular in Bologna and Padova.\par
Around 1952 Donald Glaser invented the bubble chamber (BC), may be after 
looking at the bubbles in a glass of beer [2].\par

\begin{figure}[hb]
\begin{center}
{\centering\resizebox*{!}{7 cm}{\includegraphics{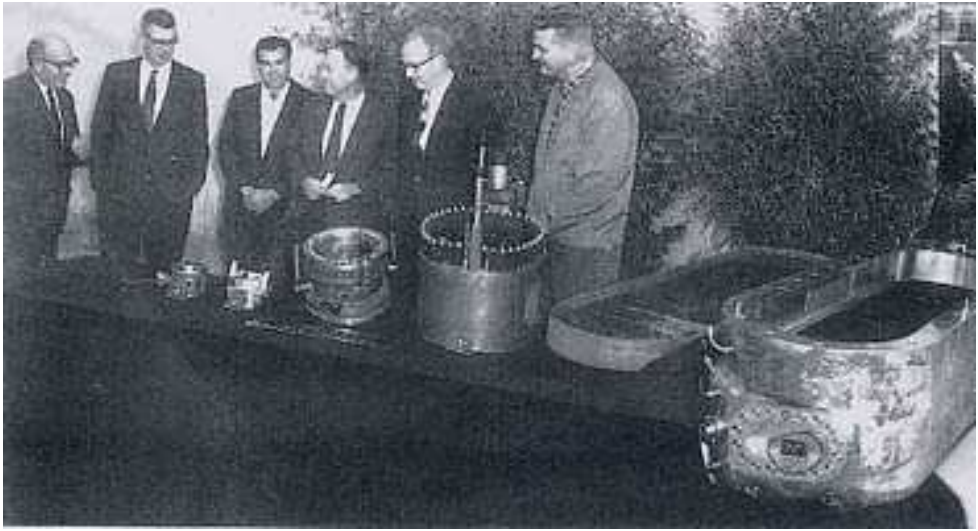}}\par}
\begin{quote}
\caption {The 2", 4", 6", 10", 15" and 72" bubble chambers
 built at Berkeley, in the early days of the BC era.}
\end{quote}
\label{fig1}
\end{center}
\end{figure}
 
The first bubble chambers were very small, but soon after much larger bubble 
chambers, filled with different liquids, were built [the increase in volume 
was about a million times; the largest chambers contained 40 $m^3$ of liquid].
Fig. 1 shows several Berkeley bubble chambers of increasing size. In the small 
exhibit, placed at the entrance of the Academy building, we had an early 
propane bubble chamber built by the late Piero Bassi in Padova.\par
Bubble chambers are $4\pi$ detectors of high density, transparent liquid 
material.

\begin{figure}[h]
\begin{center}
\begin{quote}
{\centering\resizebox*{!}{13.5 cm}{\includegraphics{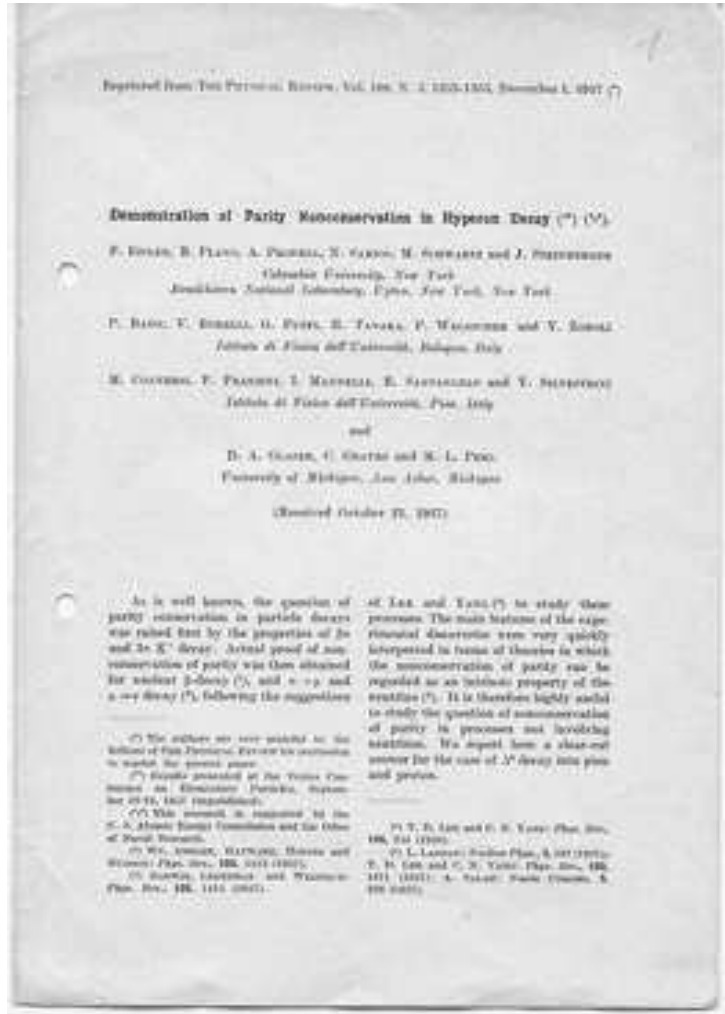}}\par}
\caption {One of the early papers published in 1957 by a
  Columbia, BNL, Bologna, Pisa, Michigan Collaboration.}
\end{quote}
\label{fig2}
\end{center}
\end{figure}

More than 100 bubble chambers were built and over 100 million stereo 
photographs were taken. While the first  BCs took small numbers of pictures 
the latest chambers took millions of photographs.\par
It turned out that bubble chambers are very reliable detectors and that they 
could be easily used in beams of unseparated and separated charged particles 
of increasing energies. Long time ago our estimeed colleague Charles Peyrou 
said in his inimitable style: ``Cloud chambers are like ladies, very delicate 
to handle, but bubble chambers are like prostitutes: they will work for 
anyone".\par
Bubble chambers needed low intensity  high quality beams of charged particles 
(about 10 particles per pulse); in many cases high voltage separators were 
used to obtain pure beams of charged K mesons. The refined high energy muon 
neutrino beams needed all the proton intensity they could get.\par
 The list of discoveries and of the studies performed over 30 years 
with the bubble chamber technique is impressive: new particles and new 
resonances, high energy hadron interactions, neutrino interactions, new 
particle searches, ... Fig. 2 shows one of the earliest papers [3]: 
``Demonstration of parity non conservation in hyperon decay": the experiment 
was performed by 4 teams from the US and Italy, [20 physicists in total]: 
the Columbia-BNL team was headed by Jack Steinberger, the Bologna team by 
Gianni Puppi and the late Piero Bassi, the Pisa team by the late Marcello 
Conversi and the Michigan team by Donald Glaser, the inventor of the bubble 
chamber.\par
Another early experiment used a liquid helium BC built at Duke University: 
Fig. 3 is a photograph of two important physicists, Martin Block and Gianni 
Puppi: notice how young they are!\par 
The sociology of bubble chamber collaborations was an interesting one. 
After an initial period when many small chambers were built almost everywhere, 
medium size BCs took relatively few pictures which were mainly analyzed by the 
``in-house" groups; later the new larger bubble chambers were built and run by 
groups of experts in large laboratories, using refined beams at increasing 
energy accelerators. These chambers were considered facilities  to be used by 
internal and external groups. This started the international collaborations, 
usually with several groups from different countries and a total number of 
physicists between 20 and 50. But the role of large laboratories, like CERN 
and Fermilab, was always a central role.\par 

\begin{figure}[h!]
\begin{center}
\begin{quote}
{\centering\resizebox*{!}{7 cm}{\includegraphics{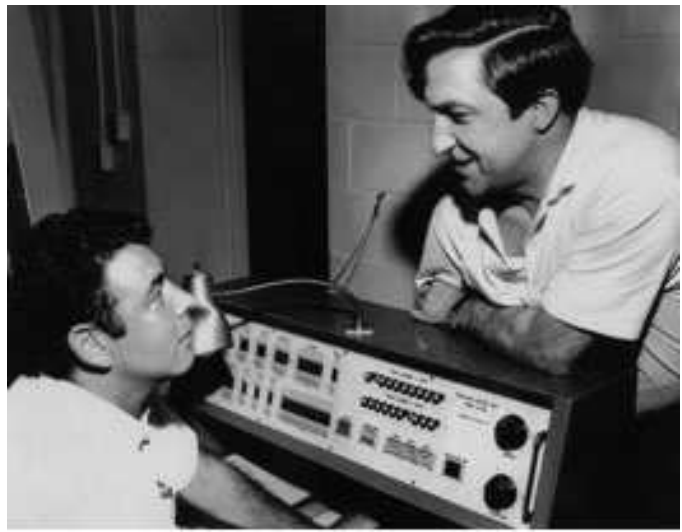}}\par}
\caption {Photograph of Martin Block and Giampietro Puppi at
  the end of the 1950's when they were collaborating in experiments
  with the Duke helium bubble chamber.}
\label{fig3}
\end{quote}
\end{center}
\end{figure}

The early bubble chamber era was dominated by U.S. groups, Berkeley and the 
Alvarez group in particular. CERN arrived later, but the 80 cm Saclay bubble 
chamber, the 2m BC and BEBC took an incredible number of pictures, analyzed in 
very many European and non European Universities.\par 
In Italy the bubble chamber technique lead to a revival of fundamental 
research in particle physics, with proficuous cooperations between 
Departments of Physics and Sections of INFN (the National Institute for 
Nuclear and Subnuclear Physics). Later the CNAF-INFN center in Bologna played 
a central coordinating role for precise measurements and for central 
computing. It may be worthwhile to recall that every team started scanning 
and measuring bubble chamber photographs with almost primitive equipment, 
like the template that you may see in our small exhibition.\par

\begin{figure}[h!]
\begin{center}
\begin{quote}
{\centering\resizebox*{!}{6.5 cm}{\includegraphics{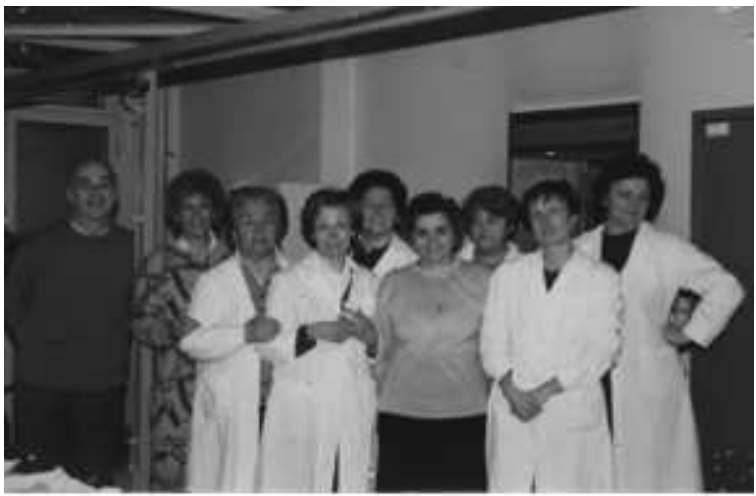}}\par}
\caption {A group of bubble chamber scanners (osservatori
  analisti) in Bologna.}
\label{fig4}
\end{quote}
\end{center}
\end{figure}

Later digitized tables were constructed and one started to hear of 
``Mangiaspagos" in Italy, of more elaborated semiautomated or fully automated 
``Frankensteins" and PEPRs in the U.S., MYLADYs and HPDs in Europe.\par
A large number of Scanners (called Osservatori Analisti in Italy) was needed 
to cope with the increasing number of photographs, with more precise 
measurements and of pre-measurements. Most scanners were pretty  young girls, 
like those in Fig. 4.\par
Computing. At the very beginning of the BC era the Slide Rule was the ``most 
powerful computer" (Fig. 5) and soon after electromechanical calculators 
(Fig. 6). But soon the IBM650 computer and, immediately after, more and more 
powerful computers started to be heavily used. The measured coordinates of 
points along the tracks were initially punched on cards manually, later 
automatically by semiautomatic projectors. The installation of mainframe  
computing capacity was driven by the demands of bubble chamber physics: the 
CERN mainframe central computers increased their capacity and speed by over a 
factor of 1000 during the bubble chamber era.\par
Various types of specialized bubble chambers were also built. Rapid cycling 
BCs, high resolution BCs, holographic BCs. Prof Gigli from Pavia built his 
``soda pop" bubble chamber [``camera a gazosa"] which behaved quite nicely. 
The latest and largest BCs needed electronic detectors, like the External 
Muon Identifier for BEBC at CERN (Fig. 7) and for the 15 foot BC at 
Fermilab.\par
The bubble chambers lead to the discovery of very many resonances, strange and 
non-strange [2]; together with the results of simple and complex electronic 
experiments [4, 5] they cleared the situation of the hadron spectrum. Bubble 
chambers were $4\pi$ detectors. Instead most of the early electronic 
experiments 
covered small solid angles, but progressively they grew bigger, covered larger 
solid angles and eventually became $4\pi$ detectors, like all the LEP 
detectors 
[6] and the LHC detectors [7]. The bubble chamber experiments initiated on a 
relatively large scale the symbiosis between large laboratories, like CERN and 
Fermilab, and their community of users.\par

\begin{figure}[h!]
\begin{center}
\begin{quote}
{\centering\resizebox*{!}{12 cm}{\includegraphics{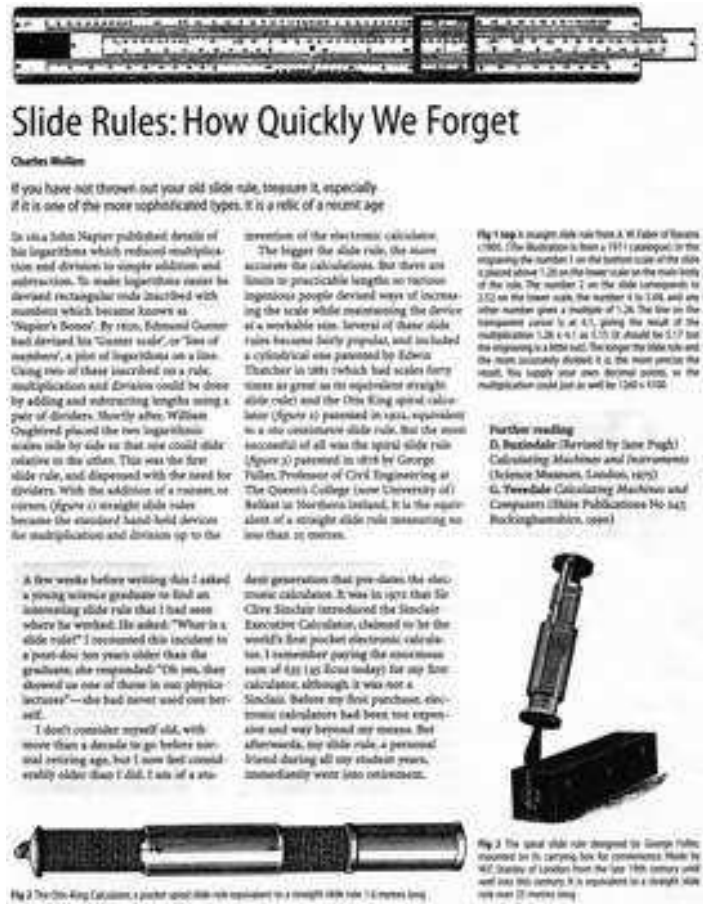}}\par}
\caption {The Slide Rule.}
\label{fig5}
\end{quote}
\end{center}
\end{figure}

\begin{figure}[h!]
\begin{center}
\begin{quote}
{\centering\resizebox*{!}{6 cm}{\includegraphics{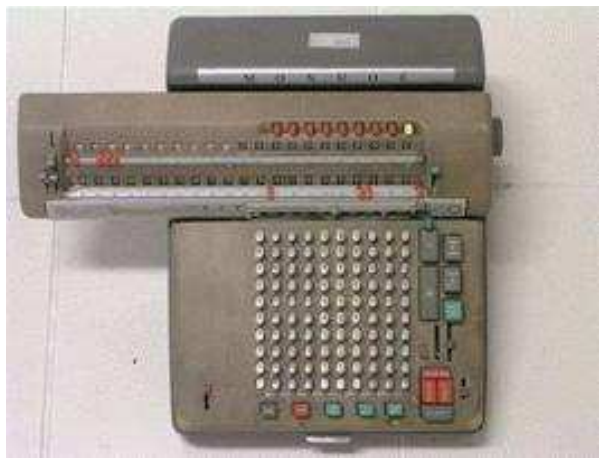}}\par}
\caption {A "Monroe" electromechanical desk calculator.}
\label{fig6}
\end{quote}
\end{center}
\end{figure}

The main scientific legacy of the bubble chamber to our understanding of the 
microworld of particle physics can probably be summarized as follows [1] [2] 
[8]:\\
- Discovery of several strange particles, like the Omega minus.\\
- Discovery of many meson and baryon resonances which lead to the knowledge of 
the hadron spectrum $\rightarrow$ SU(3) $\rightarrow$ constituent quarks.\\
- Discovery of the neutral weak current $\rightarrow$ electroweak unification.\\
- Scaling in neutrino-nucleon deep inelastic scattering $\rightarrow$ 
partons $\rightarrow$ dynamical  
quarks.\\
- Neutrino-nucleon scaling violations $\rightarrow$ support to QCD.\par

\begin{figure}[h!]
\begin{center}
{\centering\resizebox*{!}{7.4cm}{\includegraphics{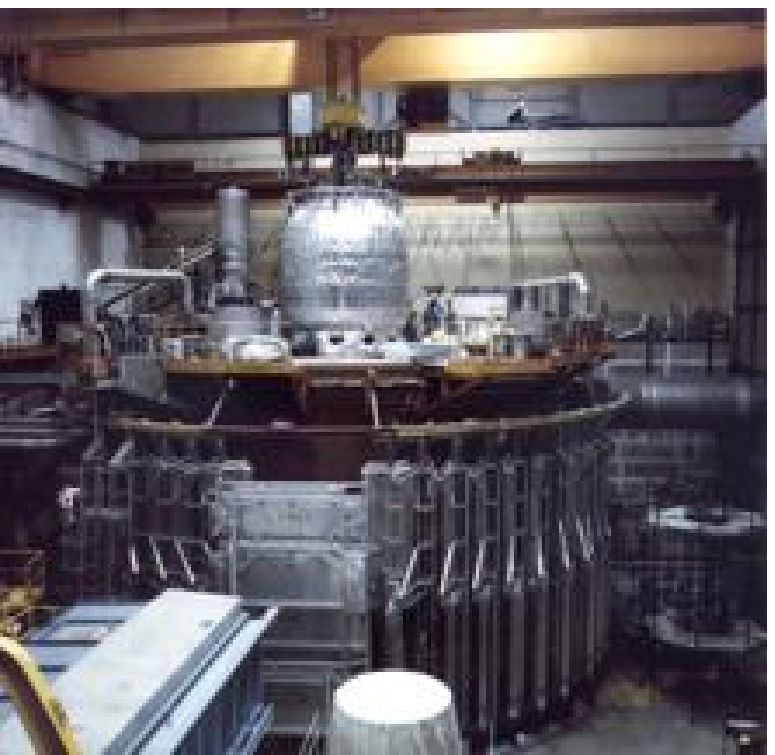}\par}}
{\centering\resizebox*{!}{7.55cm}{\includegraphics{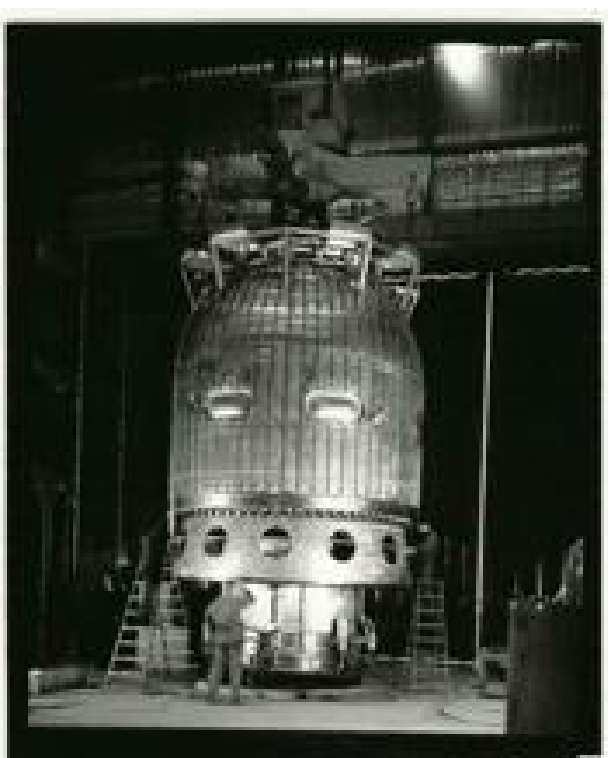}\par}}
\begin{quote}
\caption {The Big European Bubble Chamber (BEBC): the 3.7 m
  liquid container is 
shown at the top of the left figure, while it is being lowered into place; the 
right picture shows the body and in particular the expansion piston at the 
bottom. On the left, BEBC is shown with the external muon identifier composed 
of a set of electronic detectors. The superconducting coil is before the 
electronic detectors. BEBC, filled with $H_2$, $D_2$ or H/Neon was used 
mainly with 
neutrino beams.}
\label{fig7}
\end{quote}
\end{center}
\end{figure}

Popularization of High Energy Physics and public understanding of science 
started to be an important part of physics: selected bubble chamber 
photographs provide a global and intuitive view of particle physics 
phenomena; an unprepared audience and young people start to realize 
immediately that our field is based on simple and intelligible experimental 
facts [8] [9] [10]. Selected bubble chamber pictures (like those in Figs. 8, 
9, 10, 11) can also be used at the beginning  of courses on particle 
physics.\par 
The bubble chamber era lead to a large number of publications in many refereed 
scientific Journals. The publications from the various Bologna bubble chamber 
groups are quoted in [3] [11-95]. \par~\par

	I would like to thank the members of the Advisory and Organizing 
Committees, the Bologna Academy of Sciences, the Secretariats, the INFN 
Multimedia group, the colleagues who participated to the workshop and those 
who sent notes and recollections, and many other people. In particular I thank 
Dr. A. Casoni and Dr. M. Giorgini for their invaluable collaboration.

\begin{figure}[h!]
\begin{center}
{\centering\resizebox*{!}{7 cm}{\includegraphics{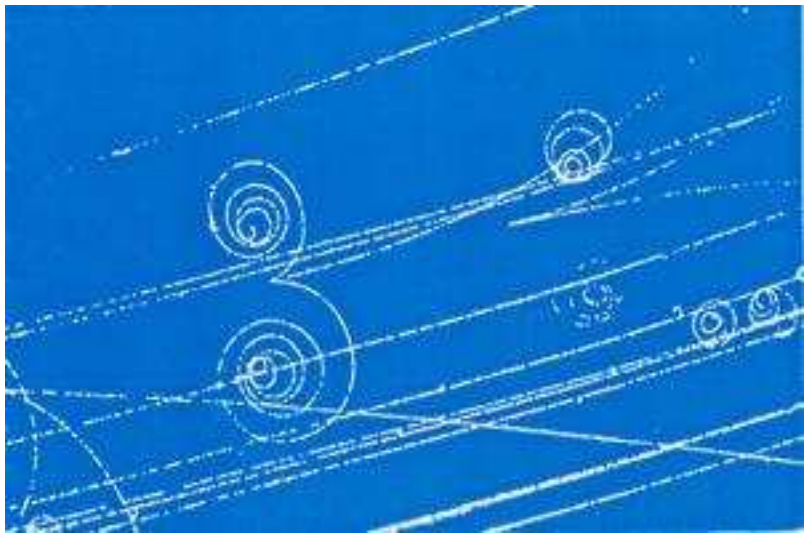}}\par}
\caption {A bubble chamber event: creation of a
  positron-electron pair by a 
photon in the Coulomb field of an electron; the electron emits, by 
bremsstrahlung, a $\gamma$ ray which eventually converts into a 
positron-electron 
pair in the Coulomb field of a proton [4].}
\label{fig8}
\end{center}
\end{figure}

\begin{figure}[h!]
\begin{center}
{\centering\resizebox*{!}{5.5 cm}{\includegraphics{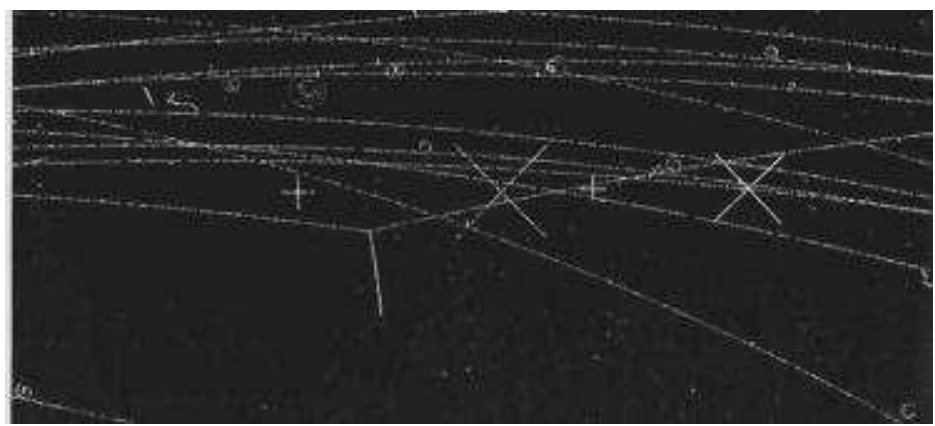}}\par}
\caption {A low energy $K^+$meson scatters elastically, $K^+ p \rightarrow 
K^+ p$, at the center of 
the photo: the $K^+$ incoming from the left scatters elastically from a proton 
of the hydrogen bubble chamber; the scattered $K^+$ goes forward, while the 
hit proton recoils at about $90^{\circ}$ and has a small energy, so that it 
stops in the chamber (BGRT Collaboration).}
\label{fig9}
\end{center}
\end{figure}

\begin{figure}[h!]
\begin{center}
{\centering\resizebox*{!}{7 cm}{\includegraphics{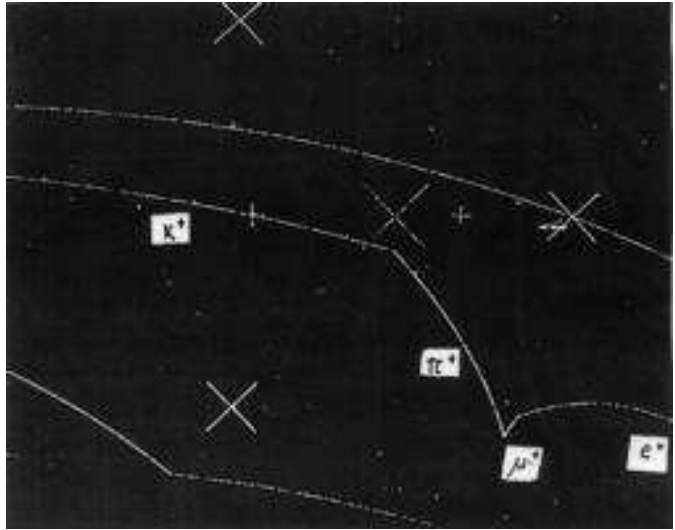}}\par}
\caption {A decay chain. A low energy $K^+$ incoming from the
  left decays in the 
center of the photograph, $K^+ \rightarrow \pi^+ \pi^{\circ}$; 
the $\pi^+$ stops and decays, $\pi^+ \rightarrow \mu^+ \nu_{mu}$; finally 
also the $\mu^+$ decays, $\mu^+ \rightarrow e^+ \nu_{e} \overline{\nu_{\mu}}$
(BGRT Collaboration).}
\label{fig10}
\end{center}
\end{figure}

\begin{figure}[h!]
\begin{center}
{\centering\resizebox*{!}{7 cm}{\includegraphics{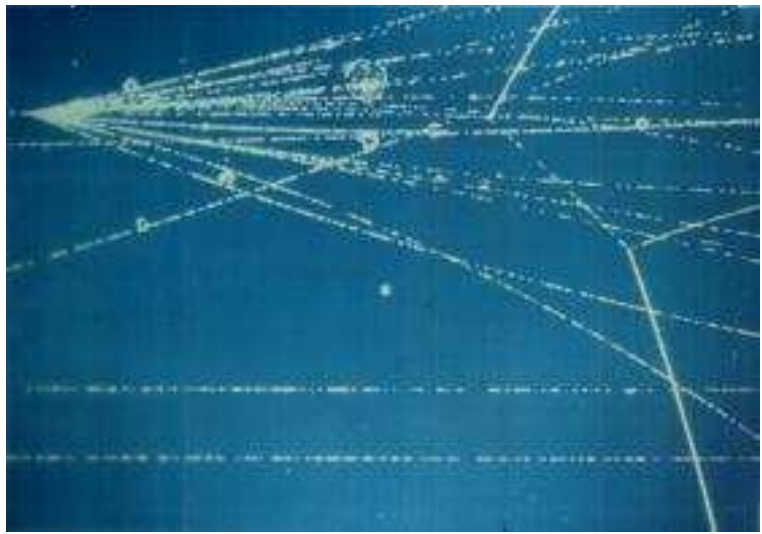}}\par}
\caption {A multiprong charged hadron production event
  produced by an incoming high energy proton (CERN photo).}
\label{fig11}
\end{center}
\end{figure}

\end{document}